\documentclass[conference]{IEEEtran}
 
\UseRawInputEncoding
\usepackage{cite}
\usepackage{amsmath}
\usepackage{amsfonts}
\usepackage{graphicx}
\usepackage{float}
\usepackage[table]{xcolor}
\usepackage{subfigure}
\usepackage{epsfig}
\usepackage{amsthm}
\usepackage{amsmath}
\usepackage{mathrsfs}
\usepackage{amsfonts}
\usepackage{amssymb}
\usepackage{mathrsfs}
\usepackage{euscript}
\usepackage{multirow}
\usepackage{url}
\usepackage{algorithm}
\usepackage{algorithmic}
\usepackage{bm}
\usepackage{booktabs}
\usepackage{placeins}
\usepackage{dirtytalk}
\usepackage{fancybox} 
\DeclareMathOperator*{\argmax}{arg\,max}

\usepackage[absolute,overlay]{textpos}

\usepackage{tcolorbox} 
 
\usepackage{xcolor, soul}
\sethlcolor{green}

\theoremstyle{remark}

\newtheorem*{remark*}{Remark}

\begin{document}

\begin{textblock*}{\textwidth}(20mm, 3mm) 
    \begin{tcolorbox}[colback=white, colframe=red, coltext=red, boxrule=0.5mm, width=\textwidth, sharp corners]
        \centering
        \textbf{\small This work has been submitted to the IEEE for possible publication. Copyright may be transferred without notice, after which this version may no longer be accessible.}
    \end{tcolorbox}
\end{textblock*}

\bstctlcite{IEEEexample:BSTcontrol}

\title{Deep Learning based Computer-vision for Enhanced Beamforming}
 
\author{Sachira Karunasena$^*$, Erfan Khordad$^*$,  Thomas Drummond$^\dagger$, Rajitha Senanayake$^*$\\
$^*$ Department of Electrical and Electronic Engineering, University of Melbourne, Melbourne, Australia\\
$^\dagger$ School of Computing and Information Systems, University of Melbourne, Melbourne, Australia\\
}

\maketitle

\begin{abstract}
 Meeting the high data rate demands of modern applications necessitates the utilization of high-frequency spectrum bands, including millimeter-wave and sub-terahertz bands. However, these frequencies require precise alignment of narrow communication beams between transmitters and receivers, typically resulting in significant beam training overhead. This paper introduces a novel end-to-end vision-aided beamforming framework that utilizes images to predict optimal beams while considering geometric adjustments to reduce overhead. Our model demonstrates robust adaptability to dynamic environments without relying on additional training data where the experimental results indicate a top-5 beam prediction accuracy of 98.96\%, significantly surpassing current state-of-the-art solutions in vision-aided beamforming.
\end{abstract}


\section{Introduction}\label{Sec:Introduction} 

The rapid evolution of wireless communication systems
is driving the shift toward millimeter-wave (mmWave) and sub-terahertz (THz) spectrums to address the demand for high data throughput \cite{environment}. However, these high frequencies
present significant challenges, particularly due to the inherent propagation and penetration loses \cite{ebs}. Therefore, obtaining higher beamforming gains by employing directional narrow beams generated by large antenna arrays is crucial. As numerous narrow beams are generated by the large antenna arrays, selecting the best beam pair for the transmitter (TX) and receiver (RX) introduces beam training overhead which is specially large in highly dynamic environments where the best beam pair needs to be updated quickly \cite{survey}.

The evolution of beam selection techniques for mmWave and terahertz communication systems has progressed from Exhaustive Beam Sweeping (EBS) \cite{ebs} methods to more sophisticated approaches \cite{xiao2016hierarchical,yan2019wideband} due to their substantial beam training overhead. EBS uses predetermined analog beam codebooks for beam sweeping at both the TX and RX ends, determining the optimal beam pair through a process that measures the received signal power.
Subsequent advancements sought to reduce this overhead through various strategies such as tree search algorithms \cite{xiao2016hierarchical} and multi-resolution codebook designs \cite{yan2019wideband} which attempt to minimize the number of beams evaluated. 
Recent innovations using wide beams on the initial level to narrow the search space and then applying high-resolution techniques show promise in reducing the overhead and improving the accuracy of beam predictions in dynamic scenarios \cite{marandi2018adaptive}. 

In recent years, machine Learning (ML) solutions have  been utilized to reduce the beam training overhead and improve accuracy of beam prediction both with \cite{nocturnal, environment, kim2024computer, heath, morais2023position} and without \cite{deepfast, ma2021deep} the integration of additional sensor data such as GPS coordinates \cite{morais2023position}, LIDAR data \cite{heath}, and colored (RGB) images \cite{nocturnal, environment, kim2024computer}. 
Focusing on RGB images, approaches in \cite{nocturnal, kim2024computer} primarily leverage visual data without incorporating mmWave information for TX identification and beam prediction. This approach limits accuracy in scenarios where the transmitter varies, as relying solely on RGB data can lead to generalization issues. Similarly, \cite{environment} relies exclusively on RGB images and environmental semantics for beam prediction, neglecting the potential enhancements from using mmWave data.

In this paper we focus on RGB images and mmWave power profiles to unveil a robust end-to-end architecture that ranges from TX identification to precise beam prediction. Our approach features two innovative, Deep Learning (DL) inspired solutions. Different to the current state-of-the-art \cite{charan2023user} that relies on two separate DL and ML models, our first solution employs a single DL model for TX identification and significantly reduces latency and computational costs. More importantly, our proposed solution for TX identification excels in generality, designed to detect the TX based solely on mmWave received signal powers without depending on color or shape.
The second DL-based solution introduced in our work focuses on beam prediction by accounting for image distortions, such as the vertical vanishing point caused by camera alignment, ensuring accurate beam directionality. By integrating these adjustments, our model achieves a high degree of generalization, making it suitable for widespread application across various operational contexts.
We employed the real-world multi-modal dataset, DeepSense 6G \cite{deepsense6g}, for our experiments, training our models on Scenario 3 and testing them on Scenarios 3 and 4. These scenarios, set in nighttime conditions, pose additional challenges due to the noise associated with low-light environments, testing the robustness of our approach in adverse settings.
In summary, we designed an end-to-end framework from TX identification to beam prediction that offers the following contributions: 
\begin{itemize}
    \item A TX identification system that detects the TX in the image based on mmWave power profiles. (Section \ref{sec:tx_indetify})
    \item A TX tracking system that sustains tracking without requiring mmWave power profiles.
    (Section \ref{sec:tx_track})
    \item A geometriccally-aware beam prediction system that incorporates vertical vanishing points to improve the accuracy of predicting top-$N$ beams under varied imaging scenarios.
    (Section \ref{sec:beam_pred})
    \item An enhanced beam prediction system with reduced beam training overhead that outperforms current state-of-the-art work \cite{nocturnal, environment}. (Section \ref{sec:results})
\end{itemize}
Our dual-pronged strategy exemplifies a significant advancement in leveraging visual data for enhanced beam management in wireless networks by outperforming the current state-of-the-art methods, \cite{nocturnal, environment} by atleast 6\% in each top-1,3,5 metrics for scenarios 3 and 4 of DeepSense 6G dataset.
\section{System Model and Problem Formulation}\label{Sec:systemModel}

\subsection{System Model}
This paper considers an Orthogonal Frequency Division Multiplexing (OFDM) mmWave system where the base station (BS) employs an RGB camera and an $M$-element uniform linear array while the single mobile RX in consideration is equipped with an omnidirectional single-antenna. Let this OFDM-enabled system be equipped with $K$ subcarriers and assume the TX, to utilize an oversampled pre-defined beamforming codebook, $\mathcal{C}_r=\{\textbf{f}_q\}_{q=1}^Q$, where $Q$ represents the total number of beamforming vectors and $\textbf{f}_q \in \mathbb{C}^{M \times 1}$. Let the channel between the RX and the BS over the $k$-th subcarrier be denoted by $\textbf{h}_k \in \mathbb{C}^{M\times1}$. 
If the transmitted signal from the $k$-th subcarrier at the BS is denoted as $x_k(t)$, the downlink received signal can be expressed as,
\begin{equation}
    y_k(t) = \textbf{h}_k^T(t)\textbf{f}_q(t)x_k(t) + v_k(t),
    \label{eqn:rx_signal}
\end{equation}
where $v_k(t)$ is the receiver complex additive white Gaussian noise at time $t$ which obeys a complex Gaussian distribution with mean zero and variance $\sigma^2$ and $\textbf{f}_q(t)$ is the $q$-th beamforming vector from the predefined codebook $\mathcal{C}_r$ at time $t$. The objective of beamforming would be to find the optimal beamforming vector at time $t$, which we denote as $\textbf{f}^*(t)$, by maximizing the average received signal-to-noise-ratio (SNR) across all subcarriers, $k \in [1, K]$, for all $\{\textbf{f}_q\}_{q=1}^Q$. 
Then $\textbf{f}^*(t)$ can be obtained as,
\begin{equation}
    \textbf{f}^*(t) = \argmax_{\textbf{f}_q(t) \in \mathcal{C}_r} \frac{1}{K} \sum^{K}_{k=1} \Vert \textbf{h}_k^T(t)\textbf{f}_q(t) \Vert ^2 \mathsf{SNR},
    \label{eqn:beam_argmax}
\end{equation}
where $\mathsf{SNR} = \frac{\mathbb{E}[\Vert x_k(t) \Vert^2]}{\sigma^2}$. 
In the following, we design a novel computer-vision-aided framework to assist in finding $\textbf{f}^*(t)$.

\subsection{Problem Formulation}

Theoretically, the optimal beam for a given time can be obtained by Eq. (\ref{eqn:beam_argmax}) using channel state information (CSI). However, with the increase in the number of antenna array elements, obtaining accurate CSI for all potential beams becomes impractical. As the number of antenna elements increases in modern sub-THz and mmWave communication systems, the resulting narrower beams complicate the task of identifying the optimal beam for transmission, thereby posing challenges in maintaining low latencies in dynamic environments. In such scenarios, one could perform an exhaustive search across all possible beams, computing the SNR for each to determine the optimal beam for communication at a given time. However, as the number of possible beams increases, which is the case in mmWave and THz communications, a large beam training overhead is required to determine the optimal beam. As a result, methods that help reduce the beam search space and assist in beam prediction is of great importance.   
As a potential solution, this work focuses on utilizing RGB images, an out-of-band input source, to develop a novel ML framework that predicts a subset of $N$ beams from a codebook of $Q$ beams. This approach results in reducing the beam search space by a factor of $\frac{N}{Q}$ \cite{heath}. In the next section, we propose a novel ML framework that achieves this task by identifying and tracking the transmitter through space and time and predicting the top-$N$ beams, which are the N beams that yield the highest SNR for each instance. 

\section{Proposed Solution: Vision-aided Beamforming}
\label{Sec:solution}

\begin{figure*}[t]
    \centering
    \includegraphics[width=1\textwidth]{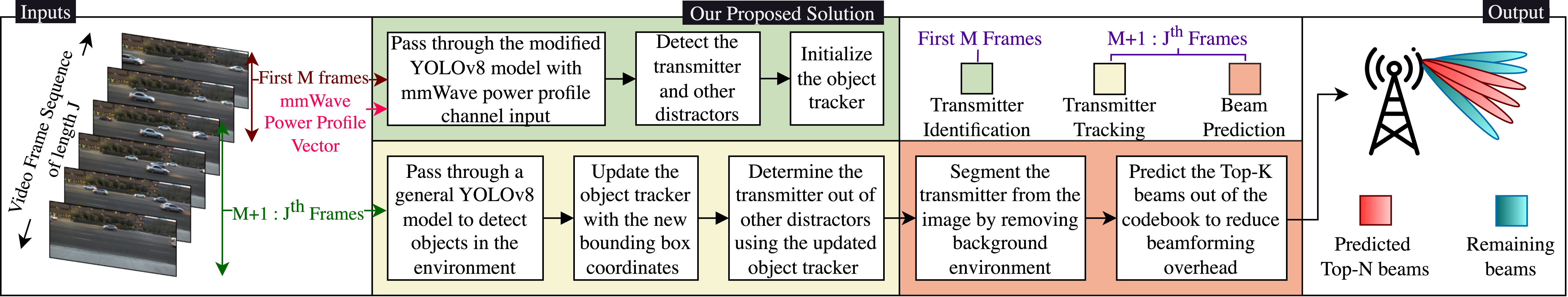}
    \caption{Proposed end-to-end Vision-aided Beamforming approach. The first $\mathbf{M}$ frames of the video sequence will be used for TX identification. The remaining $\mathbf{M+1:J^{th}}$ frame is used to track the detected TX and predict the top-$N$ beams for each frame.}
    \label{fig:framework}
\end{figure*}

This section highlights our proposed data-driven end-to-end beam selection approach guided by RGB images and mmWave received signal power profiles. Our proposed approach comprises three main sections as illustrated in Fig. \ref{fig:framework}. 
\begin{enumerate}
    \item \textbf{Transmitter Identification}: Accurately identifying the TX amongst other distractors present in the environment.
    \item \textbf{Transmitter Tracking}: Tracking the detected TX in the subsequent frames until it leaves the cell coverage area of the BS.
    \item \textbf{Beam Prediction}: Predicting the top-$N$ beams serving the tracked TX.
\end{enumerate}

\subsection{Transmitter Identification}
\label{sec:tx_indetify}

To predict the optimal beam for the TX while it remains within the BS coverage area, the first critical step is accurately identifying the TX user among multiple potential distractors. 
The images captured at the BS alone cannot determine which object is the TX.
To resolve this ambiguity, we introduce an additional mmWave power profile channel, which we append to the image input for the object detection model. 
Since mmWave beams are highly directional, the mmWave power profile provides crucial information about beam directions, indicating where the received signal power is highest which can assist in more precisely narrowing down the transmitter locations.

\begin{figure}[t]
    \centering
    \includegraphics[width=0.5\textwidth]{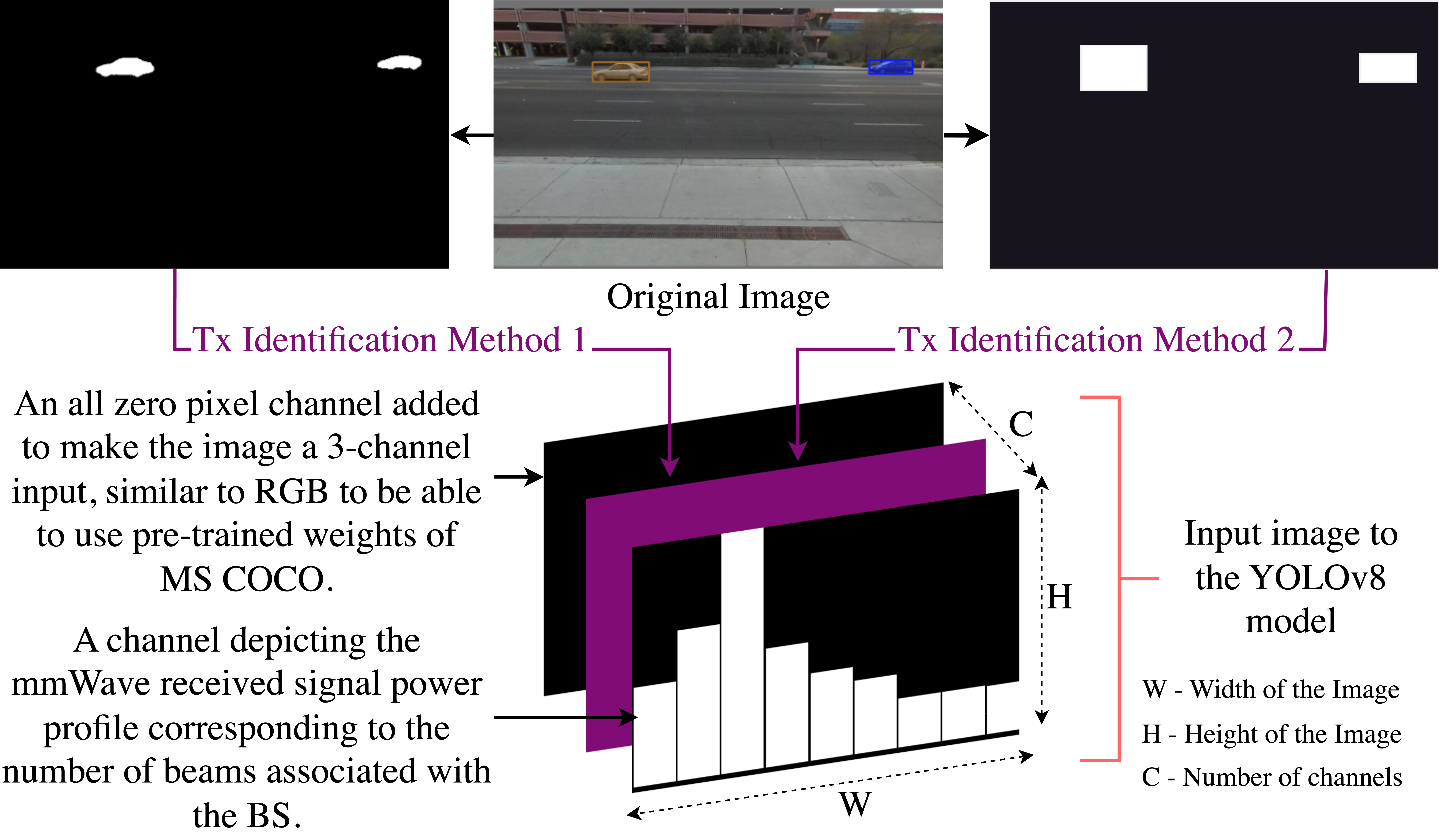}
    \caption{Structuring the input for the TX Identification methods.}
    \label{fig:tx_identi}
\end{figure}

Our proposed model for TX identification integrates the mmWave power profile directly into the object detection process, eliminating the need for a second ML model for TX identification as opposed to \cite{charan2023user} that used a second ML model. 
By incorporating the mmWave signal information into a single inference, our approach effectively reduces the latency and the number of parameters which leads to a streamline process with low computational complexity.
This is illustrated in Fig. \ref{fig:tx_identi} where the mmWave power information is added as a channel to the image, similar to how color images are composed of Red, Green and Blue (RGB) channels, where each channel represents different data dimensions in the image. We are thus building a modified 3-channel input image for the object detector.
The remaining two channels of the modified 3-channel image are configured such that 
\begin{itemize}
    \item Second Channel - mmWave Power Profile: The second channel represents each beam's received signal power profiles. The beams are aligned in a way that corresponds to the coverage area of the BS. Beams with higher received signal power indicate that the beam captures the TX object. In contrast, beams with lower signal values correspond to areas where objects, if present, are not transmitting signals to the BS.
    \item Third Channel - Zero Padding: We add an all-zero third channel to maintain compatibility with the pre-trained weights of MS COCO \cite{cocodataset}, which are based on 3-channel RGB images. 
\end{itemize}

To effectively use the mmWave power profile in conjunction with image data, we propose two methods to structure the first channel of the input image we use to train the YOLOv8 object detector \cite{YOLOv8}. In both these methodologies, we first pre-process the image by discarding the background information. 

\begin{enumerate}
    \item TX Identification Method 1: After pre-processing the image, we remove the colors of all objects, leaving only their shape. This ensures that the object detector cannot rely on color cues for TX identification.
    \item TX Identification Method 2: In this approach, we remove the color and shape of all objects within the image. The resulting input contains no visual features, and the model is compelled to depend entirely on the mmWave channel data to detect and identify the TX object.
\end{enumerate}

This 3-channel input is then fed into the YOLOv8 object detector. 
The objective of the object detector is to optimize the mapping function $g$,
\begin{equation}
    g: \{[I(t),P(t)] \rightarrow bbox_{TX\_p}(t)\},
\end{equation}
where $I(t)$ and $P(t)$ represent the image and mmWave power profile of each beam captured at time $t$ and $bbox_{TX\_p}(t)$ is the predicted bounding box (bbox) of the TX at time $t$. $bbox_{TX\_p}(t)$ contains $x, y, w, h$, which are the horizontal and vertical coordinates of the center point of the bbox and the width and height of the bbox respectively. 

To test our Tx identification process, we use Scenario 3 of the DeepSense 6G dataset \cite{deepsense6g} by splitting the data for each of the 64 beams in a 70:30 ratio for training and testing. The model is also tested on the entirety of Scenario 4 of the DeepSense 6G dataset. In summary, the model is trained on 1204 training samples randomly picked from each beam, and tested on 288 Scenario 3 and 275 Scenario 4 test samples.
The accuracy (Acc.) of this system is calculated as follows:
\begin{equation}
    \mathsf{Acc.} = \frac{\sum_D\mathbb{I}\{\mathsf{IoU}(bbox_{TX\_p}(t), bbox_{TX\_gt}(t)) \geq Z\}}{D},
\end{equation}
where $\mathbb{I}$ is the Indicator Function, $D$ is the number of samples in the test set and IoU is defined as the Intersection over Union of the two bboxes; the predicted TX, $bbox_{TX\_p}(t)$ and the ground truth TX, $bbox_{TX\_gt}(t)$ for each time instance $t$ and $Z$ is the user-defined IoU threshold.

In Scenarios 3 and 4 of the DeepSense 6G dataset, the TX object across all video frames is consistently the same car. If we were to train an object detector without running the image through the pre-processing step where we remove either the color or both color and shape, the model would inadvertently learn visual characteristics such as the shape and color of the car to identify the TX, rather than relying on the mmWave channel input. This results in a model that bases its detection on visual cues, rather than the intended mmWave received signal power profiles, essentially training a model biased for a specific scenario. As such, the existing methods described in \cite{nocturnal,kim2024computer}, do not offer a generalized solution. Our proposed methodology for TX identification addresses this by removing the visual cues and forcing the model to identify the TX based on the mmWave power profiles.

\begin{table}[t]
\caption{DeepSense 6G TX Identification Accuracy: Models trained on Scenario 3 and tested on both Scenarios 3 and 4.}
\centering
\resizebox{0.5\textwidth}{!}{%
\begin{tabular}{p{0.02cm}p{5cm}p{0.75cm}p{0.75cm}p{0.7cm}}
\toprule
 & \multirow{2}{*}{\textbf{Methodology}} & \multicolumn{3}{l}{\textbf{TX Identification Accuracy}} \\
    &   & $M$=1   & $M$=3  & $M$=5 \\
\midrule
\multirow{5}{*}{\rotatebox{90}{Scenario 3}}& Charan \textit{et al.} \cite{charan2023user} & 98.43\% & 99.00\% & 99.48\% \\
& \textbf{Ours: TX Identification Method 1} & \textbf{99.65\%} & \textbf{100.00\%} & \textbf{100.00\%}\\
& \textbf{Ours: TX Identification Method 2} & \textbf{96.52\%} & \textbf{100.00\%} & \textbf{100.00\%}\\
& Ablation study 1: RGB input & 99.10\% & 100.00\% & 100.00\% \\
& Ablation study 2: All zero mmWave channel & 24.92\% & 25.61\% & 26.82\%\\
\midrule
\multirow{5}{*}{\rotatebox{90}{Scenario 4}}& Charan \textit{et al.} \cite{charan2023user} & 97.16\% & 98.48\% & 98.60\% \\
& \textbf{Ours: TX Identification Method 1} & \textbf{97.46\%} & \textbf{99.64\%} & \textbf{99.64\%}\\
& \textbf{Ours: TX Identification Method 2} & \textbf{94.20\%} & \textbf{99.64\%}& \textbf{99.64\%}\\
& Ablation study 1: RGB input & 98.91\% & 99.64\% & 99.64\%\\
& Ablation study 2: All zero mmWave channel & 22.89\% & 23.72\% & 24.45\%\\
\bottomrule
\end{tabular}%
\label{tab:tx_identi_results}
}
\end{table}

In Table \ref{tab:tx_identi_results}, we benchmark our results against the current state-of-the-art work \cite{charan2023user}, demonstrating the impact of removing visual cues and highlighting the superior performance of our object detector that uses the modified input image. We set the IoU threshold $Z=0.5$ for our analysis.
In our proposed model for TX Identification, we utilize $M$ consecutive frames, as illustrated in Fig. \ref{fig:framework}, to ensure accurate identification of the TX. This approach mitigates the risk of missed detection in early frames by allowing the model to capture the TX in subsequent frames. 
The object classified as the TX most frequently across these $M$ frames is then identified as the TX. We tested this approach with different values of $M=1,3$ and $5$ to evaluate its effectiveness in improving detection accuracy as shown in Table \ref{tab:tx_identi_results}.
Furthermore, we compared the accuracy of our two proposed methods for TX Identification against a YOLOv8 model that uses the original RGB image with both color and shape information. Additionally, we tested a model where the second channel, representing the mmWave power profile, was replaced with an all-zero channel, effectively removing the received signal power profile. The results demonstrate that without the mmWave channel input, the model cannot accurately detect the TX, as expected. The slight increase in accuracy when using the original RGB image shows that shape and color do not play a significant role in TX detection through our proposed method, emphasizing that the mmWave power profile drives accurate identification in our approach. The comparison can be seen in Table \ref{tab:tx_identi_results}.

\subsection{Transmitter Tracking}
\label{sec:tx_track}
Following the precise detection of the TX by our TX identification methods, we deploy an object tracker DeepOCSort \cite{deepOCSort} to facilitate efficient tracking. This integration avoids the need for exhaustive beam searches in subsequent frames. 
The tracking process is supported by a three-stage system as illustrated in Figure 1. Initially, the general YOLOv8 \cite{YOLOv8} model detects objects in the environment. 
The object tracker then updates with the new bounding box coordinates 
and the labels of each object in subsequent frames, distinguishing between the TX and potential distractors. 
Furthermore, once the TX is identified, we continue to track it without masking its color or shape. As the TX remains constant within the BS coverage area, 
this approach ensures that the tracker can leverage visual cues to consistently recognize the TX's identity, avoiding any potential errors in identification.

The tracker updates the labels for each object's new bbox coordinates based on its history and visual data comparison between its previous frames. With the updated bbox coordinates of the TX, the final stage of our proposed approach uses this information to perform accurate beam prediction, ensuring efficient beam alignment.

\subsection{Beam Prediction}
\label{sec:beam_pred}
\begin{figure*}[!t]
    \centering
    \subfigure[Beam shape designs based on the codebook.]{
        \includegraphics[width=0.48\textwidth]{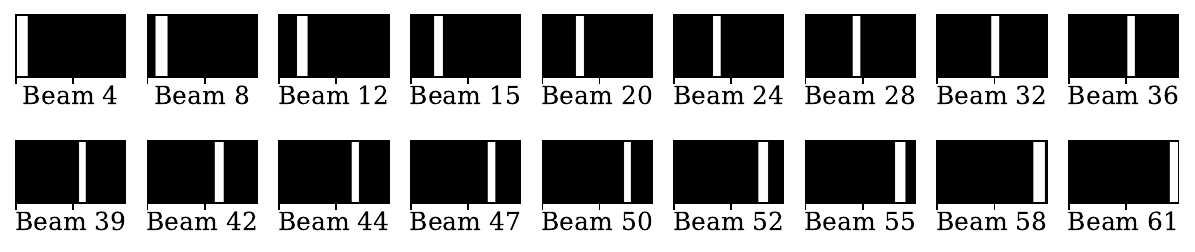}
        \label{fig:beam_width}
    }
    \hfill
    \subfigure[Beam shape designs based on the vertical vanishing point.]{
        \includegraphics[width=0.48\textwidth]{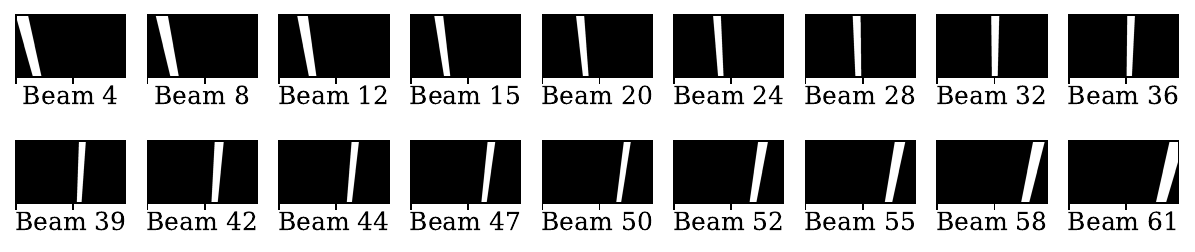}
        \label{fig:beam_width_vanishing}
    }
    \caption{Beam shape designs for Scenarios 3 \& 4 of DeepSense 6G Dataset.}
    \label{fig:mainfigure}
\end{figure*}
In this subsection, we introduce a novel two-stage approach to predict the top-$N$ beams for the identified and tracked TX, discussed in sections \ref{sec:tx_indetify} and \ref{sec:tx_track}. 
This method represents the culminating phase of our proposed solution for vision-aided beamforming.

\subsubsection{\textbf{Beam Search Space Reduction}}
Scenarios 3 and 4 of the DeepSense 6G dataset provide access to the codebook design used and, implicitly, the angular regions covered by each beam. This angular span of each beam is mapped onto corresponding portions of the image as illustrated in Fig. \ref{fig:beam_width}. This mapping reveals that while there are overlaps between subsequent beams, each beam still covers a substantial and distinct portion of the image. However, the camera used to capture images in the dataset, is positioned above street level, where the TX object traverses.
By carefully considering the images in the dataset (such as the Original image shown in Fig. \ref{fig:tx_identi}), it can bee seen that the camera is slightly tilted downwards to capture a comprehensive view of the street.
This camera tilt introduces a distortion where vertical structures within the captured images do not appear vertical and we took this into consideration when designing our beam prediction solution. 

To address this, we analyzed the images in the dataset to calculate the vertical vanishing point: The point at which the lines, which should appear vertical, converge due to the perspective introduced by the camera's tilt. Building on the vanishing point analysis, we redesigned the beam mappings to better align with the visual data presented in the image.
The redesigned beam shapes, which account for this perspective skew, are illustrated in Fig. \ref{fig:beam_width_vanishing}.
Unlike the strict vertical alignment shown in Fig. \ref{fig:beam_width}, the new beam directions stem from the calculated vanishing point. 
This projecting of the beams reflects the perspective distortions present in the images,
thereby improving the reliability and accuracy of our beam prediction model. 

Building on the vertical vanishing point and beam shape redesign, we isolate the TX within the image by removing all extraneous objects and background elements. This isolation is achieved using the bbox coordinates of the TX provided by the preceding tracking system, ensuring that only the pixels representing the TX remain in the image. Subsequently, we overlay each redesigned beam shape, adjusted for the effect of the vanishing point, onto this isolated TX image. We then identify a subset of beams from the original codebook that exhibit at least one overlap pixel with the isolated TX which is utilized in the subsequent stage of our approach.

\subsubsection{\textbf{Top-N Beam Prediction}}
\begin{figure}[t]
    \centering
    \includegraphics[width=0.5\textwidth]{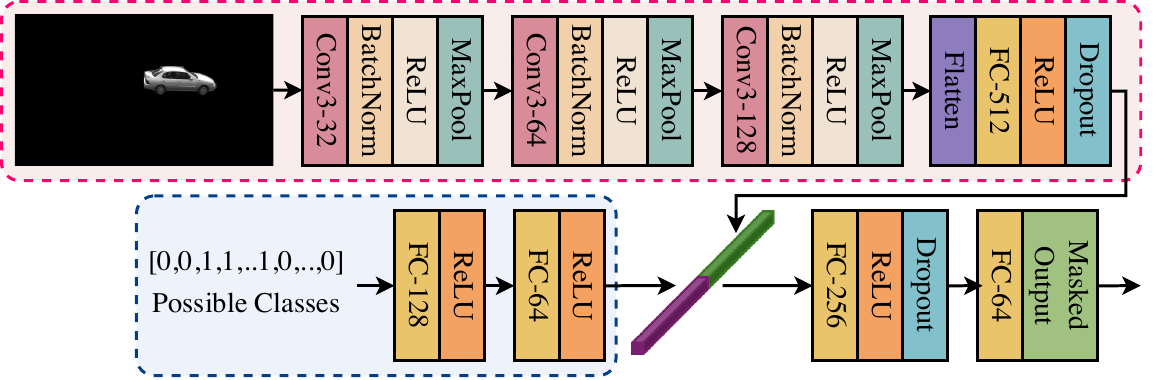}
    \caption{Custom Neural Network Architecture for Top-$N$ Beam Prediction with dual processing of Isolated TX Image and Reduced Beam Search Space.}
    \label{fig:nn}
\end{figure}

We designed a custom neural network to predict the top-$N$ beam indices for each instance of the TX. The architecture of this neural network as illustrated in Fig. \ref{fig:nn} comprises two distinct pathways:
\begin{itemize}
    \item \textbf{Image Processing: }This path processes the input image, an isolated TX image. It sequentially passes through three convolution blocks. Each block consists of a Convolutional layer (Conv), a Batch Normalization layer (BatchNorm), and a Rectified Linear Unit (ReLU) activation, followed by a Max Pooling layer (MaxPool). This sequence effectively captures and enhances the salient features of the image. Ultimately, the learned features are flattened. This flattening step converts the multidimensional feature maps into a one-dimensional feature vector, preparing it for concatenation with the second pathway.
    \item \textbf{Beam Search Space Vector Processing: }This path processes the reduced beam search space, derived from the earlier step where each possible beam index is encoded as \say{1} and all others as \say{0}. This encoding passes through two fully connected layers (FC). The features extracted from this pathway are concatenated with the flattened features from the first path.
\end{itemize}
After concatenating the feature vectors from both pathways, the combined feature vector undergoes further processing through two additional FC layers. Finally, it passes through a 64-unit classification layer, which selects the top-$N$ beam indices. Additionally, we incorporate a masking layer post-classification. This layer ensures that the output is constrained to one of the possible beams identified in the previous stage by setting the weights to zero for all other classes. 

\begin{table}[t]
\caption{Beam Prediction Accuracy: Models trained on Scenario 3 and tested on both Scenarios 3 and 4.}
\centering
\resizebox{0.5\textwidth}{!}{%
\begin{tabular}{p{0.02cm}p{5cm}p{0.75cm}p{0.75cm}p{0.7cm}}
\toprule
 & \multirow{2}{*}{\textbf{Methodology}} & \multicolumn{3}{l}{\textbf{Beam Prediction Accuracy}} \\
    &   & Top-$1$   & Top-$3$  & Top-$5$ \\
\midrule
\multirow{4}{*}{\rotatebox{90}{Scenario 3}}& \textbf{Ours} & \textbf{59.72\%} & \textbf{89.24\%} & \textbf{98.96\%} \\
& Ablation study 1: &  7.29\% & 16.67\% & 22.92\%\\
& Ablation study 2: & 56.94\%  & 87.67\% & 98.26\% \\
& Ablation study 3: & 51.39\% & 81.94\% & 93.40\% \\
\midrule
\multirow{4}{*}{\rotatebox{90}{Scenario 4}}& \textbf{Ours} & \textbf{57.83\%} & \textbf{84.42\%} & \textbf{96.01\%}\\
& Ablation study 1: & 6.82\% & 14.34\% & 21.98\% \\
& Ablation study 2: & 55.89\% & 82.23\% & 95.74\% \\
& Ablation study 3: & 50.45\% & 78.28\% & 91.77\%\\
\bottomrule
\end{tabular}%
\label{tab:beam_pred_results}
}
\end{table}

The results of our proposed beam prediction stage are presented in Table \ref{tab:beam_pred_results}. To underscore the significance of each component within our novel approach, we conducted three ablation studies:
\begin{itemize}
    \item First Ablation Study: This study evaluates the impact of image isolation by replacing the isolated image of the TX with the original image that includes other distractors and the background. This change tests the importance of focusing solely on the TX for accurate beam prediction.
    \item Second Ablation Study: In this study, we remove the second pathway, the reduced beam search space input, and the associated masked output layer, which is also the input for the second pathway. This ablation assesses the effectiveness of reducing the beam search space and masking outputs for the Top-$N$ beam prediction.
    \item Third Ablation Study: This study substitutes the redesigned beam shapes which account for the vertical vanishing point with the original beam shapes. This allows us to determine how critical the beam shape redesign is to the effectiveness of beam search space reduction.
\end{itemize}
The results of the ablation studies are presented in Table \ref{tab:beam_pred_results} where we can observe that each detail within our method significantly contributes to achieving high accuracy. This validation proves the importance of our novel approach in optimizing beam prediction, highlighting the synergistic effect of combining image isolation, beam search space reduction, and strategic beam reshaping.
We trained our neural network using cross-entropy loss, the Adam optimizer with a learning rate of 0.001, over 30 epochs, and applied a learning rate decay of 0.0001 after 59\% validation accuracy.
\section{Numerical Results}
\label{sec:results}
\begin{figure}[!t]
    \centering
    \subfigure[Top-N Beam Prediction criteria for Scenario 3]{
        \includegraphics[width=0.48\textwidth]{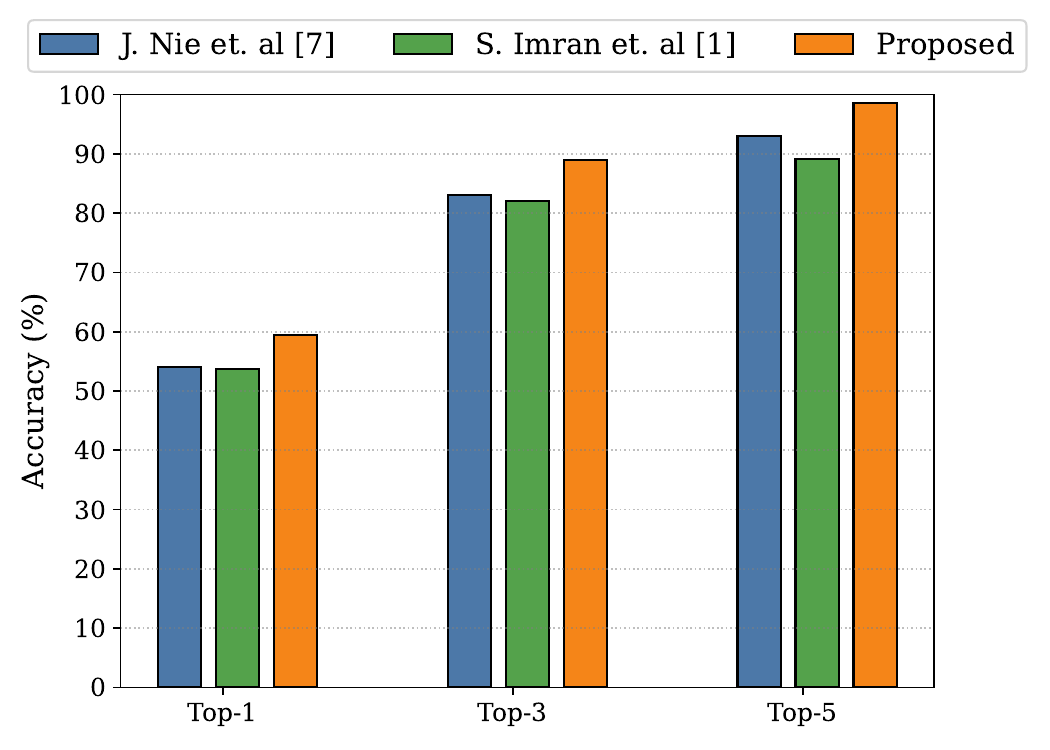}
        \label{fig:results-scne3}
    }
    \hfill
    \subfigure[Top-N Beam Prediction criteria for Scenario 4]{
        \includegraphics[width=0.45\textwidth]{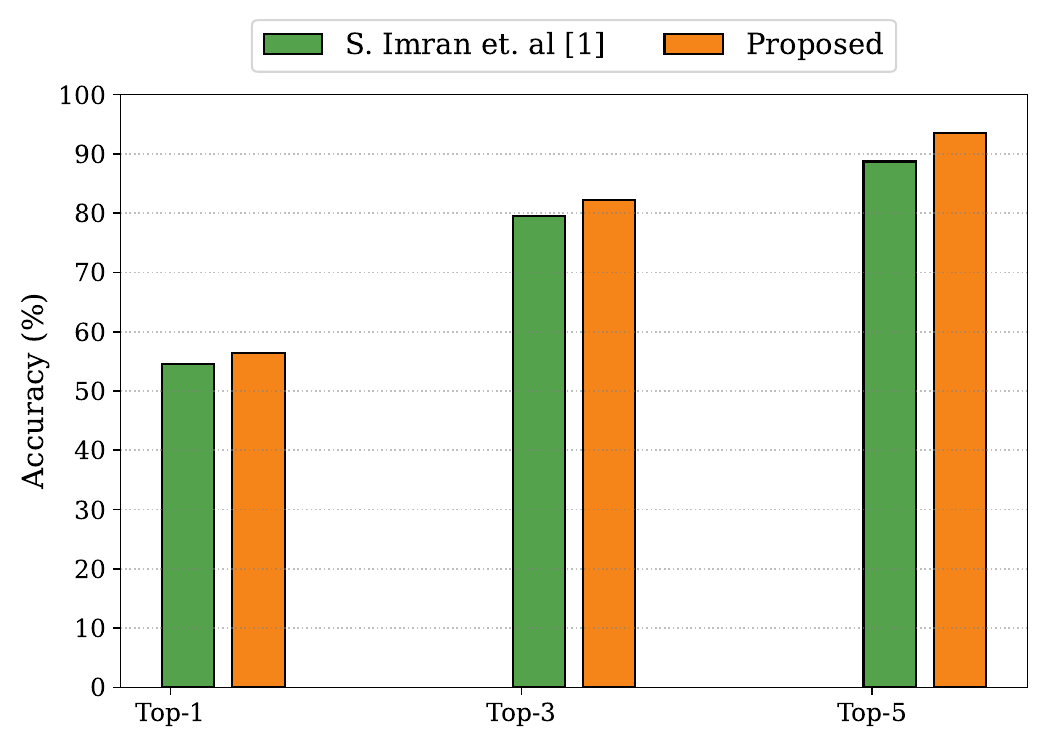}
        \label{fig:results-scen4}
    }
    \caption{Comparison of Top-N Beam Prediction metrics with current state-of-the-art methods. ``J. Nie et. al'' represents the work done in \cite{nocturnal} and ``S. Imran et. al'' represents the results generated by applying the method in \cite{environment}.}
    \label{fig:results}
\end{figure}

In preceding sections \ref{sec:tx_indetify}, and \ref{sec:beam_pred}, we detailed the individual performances of TX Identification and Beam Prediction processors. This section synthesizes the collective performance of these components and benchmarks our overall approach against current state-of-the-art. For the numerical results in this section, we use the same parameters as described in Section \ref{sec:tx_indetify}.
In Fig. \ref{fig:results}, we illustrate the overall performance of our proposed approach which exclusively utilizes vision data, in comparison to \cite{environment} and hybrid vision-position methods from \cite{nocturnal} that incorporate additional sensing information like GPS data.
Fig. \ref{fig:results-scne3} which is plotted for Scenario 3 shows that our proposed framework achieves 59.51\%, 88.93\% and 98.61\% accuracies for Top-1, Top-3, and Top- 5 metrics, as defined in Section \ref{Sec:systemModel}. Fig. \ref{fig:results-scen4} which is plotted for Scenario 4 shows that our proposed framework achieves 56.36\%, 82.27\% and 93.57\% accuracies for the same metrics.
Our integrated approach consistently surpasses existing methodologies by at least 6\% across the Top-1, Top-3, and Top-5 beam prediction metrics.
Notably, our methodology achieves comparable accuracies to the combined approach in \cite{nocturnal}, emphasizing the effectiveness of our technique without data from sensors.

\section{Conclusion}\label{Sec:conclusion}
This paper presents a comprehensive end-to-end solution for TX identification, TX tracking, and beam prediction that is adaptable to any dynamic communication scenario. Our unique method employs mmWave power profiles, integrated as a channel within a modified 3-channel input image, processed through a YOLOv8 object detector for TX detection. Subsequently, our system adeptly tracks the TX across subsequent frames and predicts the top-$N$ beams, finely adjusting for geographical nuances such as vertical vanishing points. Our approach significantly enhances accuracy—approaching near 100\% in top-5 beam predictions while reducing training overhead by over 90\%. The results affirm the usefulness of using RGB data as a sensory input, enabling near-perfect beam prediction across varied conditions with minimal overhead.


\begin{thebibliography}{10}
\providecommand{\url}[1]{#1}
\csname url@samestyle\endcsname
\providecommand{\newblock}{\relax}
\providecommand{\bibinfo}[2]{#2}
\providecommand{\BIBentrySTDinterwordspacing}{\spaceskip=0pt\relax}
\providecommand{\BIBentryALTinterwordstretchfactor}{4}
\providecommand{\BIBentryALTinterwordspacing}{\spaceskip=\fontdimen2\font plus
\BIBentryALTinterwordstretchfactor\fontdimen3\font minus \fontdimen4\font\relax}
\providecommand{\BIBforeignlanguage}[2]{{%
\expandafter\ifx\csname l@#1\endcsname\relax
\typeout{** WARNING: IEEEtran.bst: No hyphenation pattern has been}%
\typeout{** loaded for the language `#1'. Using the pattern for}%
\typeout{** the default language instead.}%
\else
\language=\csname l@#1\endcsname
\fi
#2}}
\providecommand{\BIBdecl}{\relax}
\BIBdecl

\bibitem{environment}
S.~Imran \emph{et~al.}, ``Environment semantic aided communication: A real world demonstration for beam prediction,'' in \emph{2023 IEEE International Conference on Communications Workshops (ICC Workshops)}.\hskip 1em plus 0.5em minus 0.4em\relax IEEE, 2023, pp. 48--53.

\bibitem{ebs}
Y.-N.~R. Li \emph{et~al.}, ``Beam management in millimeter-wave communications for {5G} and beyond,'' \emph{IEEE Access}, vol.~8, pp. 13\,282--13\,293, 2020.

\bibitem{survey}
M.~Q. Khan \emph{et~al.}, ``Machine learning for millimeter wave and terahertz beam management: A survey and open challenges,'' \emph{IEEE Access}, vol.~11, pp. 11\,880--11\,902, 2023.

\bibitem{xiao2016hierarchical}
Z.~Xiao \emph{et~al.}, ``Hierarchical codebook design for beamforming training in millimeter-wave communication,'' \emph{IEEE Transactions on Wireless Communications}, vol.~15, no.~5, pp. 3380--3392, 2016.

\bibitem{yan2019wideband}
H.~Yan \emph{et~al.}, ``Wideband millimeter-wave beam training with true-time-delay array architecture,'' in \emph{2019 53rd Asilomar Conference on Signals, Systems, and Computers}.\hskip 1em plus 0.5em minus 0.4em\relax IEEE, 2019, pp. 1447--1452.

\bibitem{marandi2018adaptive}
M.~K. Marandi \emph{et~al.}, ``An adaptive sequential competition test for beam selection in massive {MIMO} systems,'' in \emph{2018 IEEE 10th Sensor Array and Multichannel Signal Processing Workshop (SAM)}.\hskip 1em plus 0.5em minus 0.4em\relax IEEE, 2018, pp. 104--108.

\bibitem{nocturnal}
J.~Nie \emph{et~al.}, ``An efficient nocturnal scenarios beamforming based on multi-modal enhanced by object detection,'' in \emph{2023 IEEE Globecom Workshops (GC Wkshps)}.\hskip 1em plus 0.5em minus 0.4em\relax IEEE, 2023, pp. 515--520.

\bibitem{kim2024computer}
S.~Kim \emph{et~al.}, ``Computer vision-aided beamforming for {6G} wireless communications: Dataset and training perspective,'' in \emph{ICC 2024-IEEE International Conference on Communications}.\hskip 1em plus 0.5em minus 0.4em\relax IEEE, 2024, pp. 672--677.

\bibitem{heath}
A.~Klautau \emph{et~al.}, ``{LIDAR} data for deep learning-based mmwave beam-selection,'' \emph{IEEE Wireless Communications Letters}, vol.~8, no.~3, pp. 909--912, 2019.

\bibitem{morais2023position}
J.~Morais \emph{et~al.}, ``Position-aided beam prediction in the real world: {How} useful {GPS} locations actually are?'' in \emph{ICC 2023-IEEE International Conference on Communications}.\hskip 1em plus 0.5em minus 0.4em\relax IEEE, 2023, pp. 1824--1829.

\bibitem{deepfast}
T.~S. Cousik \emph{et~al.}, ``Deep learning for fast and reliable initial access in {AI-driven 6G} mm-wave networks,'' \emph{IEEE Transactions on Network Science and Engineering}, 2022.

\bibitem{ma2021deep}
K.~Ma \emph{et~al.}, ``Deep learning assisted calibrated beam training for millimeter-wave communication systems,'' \emph{IEEE Transactions on Communications}, vol.~69, no.~10, pp. 6706--6721, 2021.

\bibitem{charan2023user}
G.~Charan \emph{et~al.}, ``User identification: A key enabler for multi-user vision-aided communications,'' \emph{IEEE Open Journal of the Communications Society}, 2023.

\bibitem{deepsense6g}
A.~Alkhateeb \emph{et~al.}, ``Deepsense {6G}: A large-scale real-world multi-modal sensing and communication dataset,'' \emph{IEEE Communications Magazine}, vol.~61, no.~9, pp. 122--128, 2023.

\bibitem{cocodataset}
T.-Y. Lin \emph{et~al.}, ``Microsoft coco: Common objects in context,'' in \emph{Computer Vision--ECCV 2014, Proceedings, Part V 13}.\hskip 1em plus 0.5em minus 0.4em\relax Springer, 2014, pp. 740--755.

\bibitem{YOLOv8}
\BIBentryALTinterwordspacing
G.~Jocher \emph{et~al.}, ``{Ultralytics YOLO},'' Jan. 2023. [Online]. Available: \url{https://github.com/ultralytics/ultralytics}
\BIBentrySTDinterwordspacing

\bibitem{deepOCSort}
G.~Maggiolino \emph{et~al.}, ``Deep {OC-SORT}: Multi-pedestrian tracking by adaptive re-identification,'' \emph{arXiv preprint arXiv:2302.11813}, 2023.

\end{thebibliography}


\end{document}